\documentclass[useAMS,usenatbib]{mn2e}
\usepackage{graphicx}

\begin{document}

\title{On the dynamics of clouds in the broad-line region of AGNs with an ADAF atmosphere }

\author[ F. Khajenabi]{ Fazeleh Khajenabi\thanks{E-mail:
f.khajenabi@gu.ac.ir;}\\
School of Physics, Faculty of Science, Golestan University, Gorgan 49138-15739, Iran}

\maketitle

\date{Received ______________ / Accepted _________________ }

\begin{abstract}
We investigate orbital motion of spherical, pressure-confined clouds in the broad-line region (BLR) of active galactic nuclei (AGN). The combined influence of gravity of the central object and the non-isotropic radiation of the central source are taking into account. While  most of the previous studies assume that the pressure of the intercloud gaseous component is proportional to a power-law function of the  radial coordinate, we generalize it to a case where the external pressure  depends on both the radial distance and the latitudinal angle. Our prescribed pressure profile determines the radius and the column density of BLR clouds as a function of their location. We also discuss about stability of the orbits and a condition for the existence of bound orbits is obtained. We found that BLR clouds tend to populate the equatorial regions more than other parts simply  because of the stability considerations. Although this finding is obtained for a particular pressure profile, we think, this result is valid as long as the pressure distribution of the intercloud medium decreases from  the equator to the pole.
\end{abstract}

\section{Introduction}

In the unified theory of active galactic nuclei (AGN), various components have been proposed to explain observational features of these interesting astronomical objects and a supermassive black hole at center of an ANG is believed to be the main engine for the entire system \cite[e.g.,][]{netzerbook}. The central black hole is surrounded by a gaseous component which is known as broad-line region (BLR) because of  the broad emission lines in its spectrum.   Several observational evidences suggest that the BLR of an AGN has a clumpy structure, where these high-density gaseous clouds are moving on their orbits \citep[e.g.,][]{krolik88,nenkova,schartmann2008,maiolino2010, torri,risaliti2011}. In fact, knowing the physical properties of AGNs and their components first requires measuring the mass of the central black hole. Kinematic of BLR clouds is one of the main source of information for determining the mass of the central supermassive black hole. For this reason, orbital motion of the BLR clouds has been studied by many authors during recent years \citep[e.g.,][]{netzer10,krause12,plewa}. In the early studies, gravity of the central object is considered as the dominant force which controls the orbital motion of each BLR cloud. However, BLR clouds may also feel an extra force due to the central radiation of the accretion disc around the supermassive black hole. Since gravity and radiation forces are both proportional to the inverse square of the radial distance, the mass of the central black hole is underestimated  if the radiation pressure force is neglected \citep{Marco}. Moreover, BLR clouds should have comfortably long lifetimes to allow significant effects on the emitted spectrum of AGNs. It has been argued that magnetic effects can provide a significant confinement mechanism \citep[e.g.,][]{Ress}, thought most of the previous studies generally neglect its dynamical role for simplicity \citep[also see,][]{wang2012,khajenabi2014}.

From the early works on the kinematics of a cloud in a radiation field, we can mention \cite{saslaw} and \cite{Mio}, who considered an anisotropic time-dependent radiation field and the orbital motion of a cloud investigated using a perturbative approach. But the force due to the radiation pressure can significantly modify orbital motion of the BLR clouds so that mass estimate of the central supermassibe black hole needs modification \citep[e.g.,][]{Marco}. This finding has been further examined by many authors \citep[e.g.,][]{netzer10}. In these studies, generally, it is assumed that the central radiation source is isotropic and the pressure profile of the intercloud medium is a power-law function of the radial distance. On the other hand, theoretical and observational considerations imply that a geometrically thin accretion disc should exist around the central black hole, where its radiation is highly non-isotropic. \cite{Liu2012} emphasized on the role of non-isotropic radiation for the distribution of dusty gas. Then, \cite{krause11} examined kinematics of BLR clouds subject to isotropic and non-isotropic radiation fields and obtained an analytical requirement for the stability of clouds under these conditions. In \citep{krause11}, like previous works, the BLR clouds are pressure-confined and the pressure distribution of the intercloud medium has a power-law dependence on the radial distance. Thus, column density of the clouds depends on the radial distance. As a particular case where the column density is uniform, \cite{plewa} obtained an analytical solution for the orbit of BRL clouds and a condition for the stability of orbits.

 Most of the previous analytical orbital analysis of BLR clouds are based on a few certain simplifying approximations. First of all, it is assumed that each cloud is in a pressure-confined state, where the internal pressure of the cloud is in equilibrium with ambient gas pressure and during cloud's motion it remains in its pressure-confined state. Moreover, the geometrical shape of a cloud is assumed to be spherical, for simplicity. Unfortunately, our knowledge about the true nature of the clumps is very limited. The nearest cloudy system around a black hole just recently discovered near the center of our galaxy which may give us some physical insights about these clouds \citep{Gill}. A dense cloud, known as G2, with three times the mass of Earth  is moving  on a highly eccentric orbit towards  the Galaxy center. Probably it has been formed as a result of captured, shock-heated stellar winds \citep{burkert2012,schart}. \cite{burkert2012} performed  numerical simulations of a cloud with the properties similar to G2 and found that despite of the variations of the cloud due to its interactions with the ambient gaseous medium which is modeled as Advection-Dominated Accretion Flows (ADAFs; Narayan \& Yi 1994), the cloud preserves its pressure equilibrium with the surrounding medium \citep[also see,][]{schart}. But according to the numerical simulations of \cite{proga} for the irradiated clouds, it seems that even dense clouds do not move as a whole because of the gravity and the radiation forces and they experience major evolution is shape and size \citep[also see,][]{nameka}. However, these simulations are for the {\it non-magnetized} case and there are notable theoretical arguments that the clouds are magnetically confined \cite[e.g.,][]{Ress}. Moreover, as we mentioned, eclipses by BLR clouds are common among AGNs and further works are needed to understand maintenance mechanism of the clumps. Thus, in our work, we follow previous analytical studies of BLR cloud's orbits by assuming they are pressure-confined and their shape is spherical, though  recent studies suggested that in a few cases these obscuring clouds may have a cometary shape \citep{maiolino2010,risaliti2011,torri}.

The main ingredient of all previous models is the true nature of the intercloud medium. Since the clouds are assumed to be pressure-confined, it is the pressure profile of the intercloud medium which has a vital role.    Because of poor knowledge about the intercloud medium, its pressure distribution has been prescribed to be uniform or as a power-law function of the radial distance.  Recently, it has been suggested by \cite{krause11}   that one of the plausible candidate to describe the intercloud medium is Advection-Dominated Accretion Flows (ADAFs), where pressure of the gas varies in proportion to a power-law function of the radial distance \citep[e.g.,][]{Narayan94}. Moreover, dynamics and  origin of clouds near to the Galactic center (like G2 cloud) have been studied based on an ADAF for describing the intercloud hot gas \citep{burkert2012,schart}.  Here, we also describe intercloud medium using an analytical model of ADAFs where pressure depends on both the radial distance and the latitudinal angle. We think this prescription is more realistic if one wants to study kinematics of BLR clouds. In the next section, we discuss about the net force on an individual BLR cloud. Then, a condition for the existence of bound orbits is obtained in section 3.  Shape of the orbits is examined in this section. We conclude with a summary of the results in section 4.

\section{Forces on Optically Thick Clouds}

For analyzing orbit of the BLR clouds in the gravitational field of a central supermassive black hole with mass $M$, we have to consider forces on each BLR cloud.  Moreover, we have to apply certain simplifying assumptions. First of all, like all previous analytical studies, we assume the clouds are pressure-confined and during their motion this state will not change. It is also assumed that the intercloud medium can be modeled using an ADAF model. Obviously, possible hydrodynamical instabilities due to the interaction of a cloud with the ambient medium are neglected for simplicity. We also assume the clouds do not interact with each other. A few authors recently  studied dynamics of an ensemble of clouds orbiting the central object analytically with the drag force \citep*[e.g.,][]{wang2012,khajenabi2014}. We can now study motion of an individual cloud with mass $m$ without the drag force instead of analyzing kinematics of an ensemble of clouds. Thus, each cloud experiences two main forces, i.e. the gravitational force and the radiation force.

The gravitational force is

\begin{equation}
{\bf F}_{\rm grav}=-\frac{GMm}{r^2} {\rm\bf e}_{\rm r},
\end{equation}
where $r$ is the radial distance and ${\rm\bf e}_{\rm r}={\bf r}/r$. But the force due to the radiation field mainly depends on the optical properties of the cloud. We assume the cloud is optically thick. Thus,
\begin{equation}
{\bf F} _{\rm rad}=\frac{\sigma}{c}\frac{L_{\rm a}|\cos \theta|}{2\pi r^2} {\rm\bf e}_{\rm r},
\end{equation}
where $\sigma$ is the cloud's cross-section and $\theta$ is polar angle. Moreover, $L_{\rm a}$ is the luminosity of the central source which is assumed to be a geometrically thin accretion disc. The term $\cos\theta$ appears because the central source of radiation is assumed to be non-isotropic as has been pointed out by \cite{Liu2012}. Therefore, the net force on a cloud becomes

\begin{equation}\label{eq:force-net}
{\bf F}(r,\theta)=\frac{GMm}{r^2}\left( \frac{3l}{\mu N_{cl}\sigma_{T} }|\cos\theta|-1\right),
\end{equation}
where $\mu$ and $\sigma_{\rm T}$ are the mean molecular weight and the Thomson cross section, respectively. Also, $l$ is the luminosity in Eddington units, i.e. $l=L_{\rm a}/L_{{\rm edd}}$, and  the Eddington luminosity  is  $L_{\rm edd}=4\pi GMm_{\rm p}c/ \sigma_{\rm T}$.

We note that when a cloud is moving on its orbit within a gaseous medium, it may lose angular momentum due to the drag force. Here, this dissipative force is neglected for simplicity. Thus, the net force is still in the radial direction, and so, the angular momentum of a cloud is conserved and its orbit would be in a plane where its orientation is determined by the initial conditions. But we know that the net force is {\it not} conservative.

The clouds should be long-lived entities because of their significant effects on the spectrum of the system. A reasonable approximation is that to assume each cloud is in a pressure-confined state, otherwise it may contract or disperse. There are strong arguments that magnetic field effects may act as a significant additional confining mechanism \cite[e.g.,][]{Ress}. At the boundary of  a cloud there would be a balance between the internal total pressure and the intercloud pressure, irrespective of the kind of possible confining mechanisms. Thus, one can calculate column density of a cloud $N_{\rm cl}$ based on this requirement. But our little knowledge about physical properties of the intercloud medium such as its density or pressure distributions raises next challenging problem. For this reason, previous authors either assume that $N_{\rm cl}$ is constant or prescribe it as a power-law function of the radial distance. It has been proposed that models for the hot accretion flows such as ADAFs can be used for describing the gaseous intercloud medium \citep[e.g.,][]{krause11}. Original ADAF solutions \citep[e.g.,][]{Narayan94} are obtained using similarity method from the height-integrated hydrodynamic equations and so,  these solutions can not adequately describe the density or pressure distributions in three dimensions. That was a good motivation for \cite{Narayan95} to re-consider ADAF models, but in a spherical system of coordinates $(r,\theta,\varphi)$ so that each physical quantity has a radial similarity profile and a $\theta$-dependence. It has been shown that height-integrated similarity solutions consistent with the solutions in the spherical system if such solutions are averaged in the $\theta$-direction. But it is not appropriate to  use the averaged solutions for calculating $N_{\rm cl}$ of a cloud which is moving on its orbit. Depending on the orbital plane inclination angle, the column density of a cloud depends on both the radial distance and the polar angle.

However, to our knowledge, all of the similarity solutions for ADAFs in the spherical coordinates are obtained by solving a set of ordinary differential equations numerically subject to the boundary conditions at the polar axis and the equator. Thus, based on these solutions, column density of a cloud is obtained numerically, and it would be difficult to use it for determining  the orbit of a cloud and discuss about its shape or condition of having bound orbits analytically. But just recently, \cite{shadmehri14} presented a set of analytical solutions for ADAFs in the spherical system of coordinates where their properties are very similar to the original solutions of \cite{Narayan95} in the spherical coordinates. These fully analytical solutions for the gaseous component is actually what we need for determining orbits of the clouds.

 Properties of ADAFs are very different from the standard accretion disc model \citep{shakura}. In this type of accretion flows which has been successfully applied to the Galactic center, the generated heat due to the turbulence does not radiate out of the system and is advected with the flow towards the central object. For this reason, ADAFs are actually radiatively  inefficient and their temperature is high in comparison to the standard disc model. Moreover, ADAFs are geometrically thick which is a direct consequence of remaining the advected energy within the system. Gradient of the pressure is generally negligible in the standard disc model, but in ADAFs this term in the momentum equation becomes significant both in the radial and the latitudinal directions. Therefore, all physical quantities of ADAFs (including pressure) depend on the radial and the polar coordinates. In self-similar models of ADAFs, the accretion rate is an input parameter and is not determined as an unknown parameter from the boundary conditions. In ADAF model which we apply, the accretion rate is an input parameter.  

By assumption, the clouds are in pressure equilibrium with the intercloud component and so, it is the spatial pressure distribution of the intercloud medium which determines the radius and the column density of each cloud. According to the standard similarity solutions for ADAFs in the spherical coordinates $(r,\theta ,\varphi )$ with the central mass $M$ at its origin,  the pressure varies in proportion to $r^{-s}$ where the power-law index for the pressure $s$ is $5/2$. This index may  take other values between 1 and 3 if outflows are also considered. As we mentioned, these theoretical considerations motivated \cite{netzer10} and \cite{plewa} to consider only the radial distribution of the pressure for analyzing dynamics of clouds. But Shadmehri (2014) presented a set of analytical solutions for the structure of an ADAF. The pressure is
\begin{equation}\label{eq:pressure}
P_{\rm gas}=P_{0}GM \frac{(\sin\theta)^{\varepsilon '}}{r^{5/2}},
\end{equation}
where
\begin{equation}
P_{0}(\varepsilon ', \alpha, \dot{m})= \sqrt{\frac{2}{\pi}} \frac{\varepsilon '}{\alpha } \frac{\Gamma (\frac{3}{2}+\frac{\varepsilon '}{2})}{\Gamma (1+\frac{\varepsilon '}{2})} \dot{m},
\end{equation}
 and $\alpha$ is the viscosity coefficient. Here, $\Gamma$ is the standard Gamma function. Moreover, we have $\epsilon ' = \epsilon / f$ and $\epsilon = (5/3 - \gamma)/(\gamma -1)$ and $f$ is the advected energy parameter and $\gamma$ is the heat capacity ratio. Also, the non-dimensional accretion rate is $\dot{m}$. Note that $P_0$ is a non-dimensional parameter which depends on the amount of the advected energy, viscosity coefficient and the non-dimensional accretion rate.

For a pressure-confined cloud, we have pressure equilibrium at the boundary of the cloud, i.e. $P_{\rm gas}=P_{\rm cl}$ where $P_{\rm cl}$ is the internal pressure of the cloud. As it has been discussed by \cite{wang2012} and \cite{khajenabi2014}, we assume that internal temperature of  each cloud  does not change significantly during its orbital motion. Thus, internal density of each cloud $\rho_{\rm cl}$ becomes in proportion to the ambient pressure, i.e. $\rho_{\rm cl} \propto P_{\rm gas}$. Assuming the mass of each cloud is conserved, we have $\rho_{\rm cl} \propto R_{\rm cl}^{-3}$ and so, $R_{\rm cl} \propto P_{\rm gas}^{-1/3}$. Since the column density through the center of a single spherical cloud $N_{\rm cl}$ is in proportion to $R_{\rm cl}^{-2}$, we have $N_{\rm cl} \propto P_{\rm gas}^{2/3}$. Therefore,
\begin{displaymath}
N_{\rm cl}\propto [P_{0}(\varepsilon ', \alpha, \dot{m})]^{2/3} (GM)^{2/3} r^{-5/3}(\sin\theta)^{2\varepsilon'/3},
\end{displaymath}
or
 \begin{equation}\label{eq:Ncl}
N_{\rm cl}=N_{0} [P_{0}(\varepsilon ', \alpha, \dot{m})]^{2/3} (r/r_0 )^{-5/3}(\sin\theta)^{2\varepsilon '/3} ,
\end{equation}
where $N_{0}$ is a constant. Thus, column density of a cloud varies depending on its orbital location. Since the force due to the radiation is inversely proportional to the column density, the net force on a cloud would have a complicated dependence on the radial distance and latitudinal angle. Upon substituting equation (\ref{eq:Ncl}) into equation (\ref{eq:force-net}), the net force becomes
\begin{displaymath}
{\bf F}(r,\theta ) =
\end{displaymath}
\begin{equation}\label{eq:Force-Final}
\frac{GMm}{r^2} \left (\frac{3l}{\mu N_0 \sigma_{\rm T}} [P_{0}(\varepsilon ', \alpha, \dot{m})]^{-2/3} (r/r_0 )^{5/3} \frac{|\cos\theta |}{(\sin\theta)^{2\varepsilon' /3}} -1 \right ).
\end{equation}

Having the above relation for the net force on a cloud, we can study orbital motion like a standard two-body problem. In the next section, we present the radial equation of motion. But before solving the orbit equation numerically, a condition for the existence of bound orbits is obtained analytically.

\section{Analysis of Orbits}
\subsection{MAIN EQUATION}
\begin{figure}
\vspace{-150pt}
\includegraphics[scale=0.65]{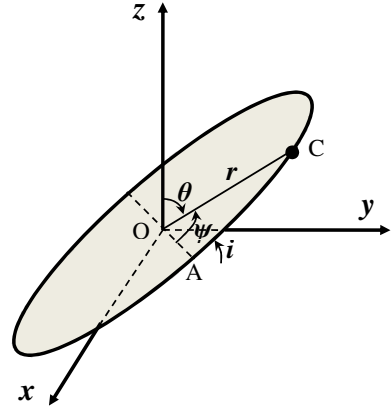}
\caption{A schematic representation of the orbital plane of a cloud.  Here, the angle $i$ denotes the inclination of the orbital plane with respect to the accretion disc. Location of the cloud is determined by $(r,\psi)$.}
\label{fig:f1}
\end{figure}

Obviously, the angular momentum of a cloud which is moving subject to the net force equation (\ref{eq:Force-Final}) is conserved, though the force is not conservative. Thus, the orbital motion would be in a plane where its orientation is determined by the initial angular momentum $L$. Orbital plane has a fixed inclination $i$ with respect to the accretion disc (i.e., equatorial plane). Figure \ref{fig:f1} shows a schematic representation of the orbital plane of a cloud.
Intersection of the orbital and the equatorial planes are shown by a dashed line. While one can denote location of a cloud by the polar angle $\theta$ and the radial distance $r$, it is more convenient to define the angle $\psi$ in the plane of motion from the ascending node (i.e., point A). Then, we have $\cos\theta = \sin(i) \sin(\psi )$. For the following analysis, we introduce these variables: $r(\psi=0)=r_{0}$ and $N_{\rm cl}(\psi=0)=N_{0} [P_{0}(\varepsilon ', \alpha, \dot{m})]^{2/3}$.

The radial equation of motion becomes
\begin{equation}\label{eq:radial-r}
 \ddot{r}-\frac{L^2}{r^3}=\frac{GM}{r^2}\left[ k (r/r_0 )^{5/3}\frac{|\sin(\psi)|}{(1-\sin(i)^2\sin(\psi)^2)^{\varepsilon '/3}}-1 \right],
\end{equation}
where $L=r^2 \dot\psi$ and
 \begin{equation}\label{eq:k}
k=k_{0}[P_{0}(\varepsilon ', \alpha, \dot{m})]^{-2/3} \sin(i),
\end{equation}
where $k_0 = 3l/ ( \mu N_0 \sigma_{T}) $.
Changing the independent variable from time $t$ to angle $\psi$ and introducing $r=1/u$, the above radial equation of motion becomes
 \begin{displaymath}
\frac{d^{2}u}{d\psi^{2}}+u = \frac{GM}{L^{2}} \times
\end{displaymath}
\begin{equation}
\left[ 1-k (u/u_{0})^{-5/3} \frac{|\sin\psi|}{(1-\sin (i)^{2}\sin (\psi)^{2})^{\varepsilon '/3}}\right ].
\end{equation}
where $u_0 = 1/r_0 $. This is the main equation of our orbit analysis which can be solved numerically subject to the appropriate boundary condition. Obviously,  shape of the orbits  depends  not only on the initial conditions, but also on the input parameters such as $k$ and the  inclination angle $i$. However, it is desirable to obtain condition of having bound orbits, because those clouds may contribute to the received emission significantly.

\subsection{CONDITION OF BOUND ORBITS}
\cite{krause11} studied stability of clouds in the BLR of an ANG for both isotropic and anisotropic light sources analytically. Their approach followed  closely the classical analysis of the two-body problem subject to a central force. For anisotropic central radiation, then, they obtained a critical column density beyond which bound orbits should be found. But in the analysis of \cite{krause11}, the pressure distribution of the gaseous component has only a radial dependence. We also follow a similar approach, but the background pressure profile has both the radial and the latitudinal dependence  according to equation (\ref{eq:pressure}).

Equation (\ref{eq:radial-r}) can be written as
\begin{equation}
\ddot{r}=\frac{GM}{r^2}\left[ k (r/r_0 )^{5/3}\frac{|\sin(\psi)|}{(1-\sin(i)^2\sin(\psi)^2)^{\varepsilon '/3}}-1 \right]+\frac{L^2}{r^3}.
\end{equation}
We introduce the ratio of the azimuthal velocity at $\psi =0$ and the Keplerian velocity at $r(\psi =0) = r_0$ by $V_0$, i.e. $V_0 = v_{\rm\psi}(\psi =0)/v_{\rm K}(r_0 ) $. Since $L=r_0 v_{\psi}(\psi =0 )$, then the angular momentum is written as
\begin{equation}\label{eq:L-r0}
L=V_0 \sqrt{GM r_0}.
\end{equation}
Thus,
\begin{displaymath}
\ddot{r}=\frac{GM}{r^2} [ k (r/r_0 )^{5/3}\frac{|\sin(\psi)|}{(1-\sin(i)^2\sin(\psi)^2)^{\varepsilon '/3}}-1
\end{displaymath}
\begin{equation}
+\frac{V_0^2}{(r/r_0 )}   ].
\end{equation}
It is as if we have one-dimensional motion in the $r$ direction subject to an effective force $F_{\rm eff}$, i.e.
\begin{displaymath}
F_{\rm eff}=\frac{GM}{r^2} \times
\end{displaymath}
\begin{equation}
\left[ k (r/r_0 )^{5/3}\frac{|\sin(\psi)|}{(1-\sin(i)^2\sin(\psi)^2)^{\varepsilon '/3}}-1 +\frac{V_0^2}{(r/r_0 )}  \right ].
\end{equation}

By analyzing this effective force, we can study properties of the orbital motion qualitatively. First, we determine the force-free locations for which the effective force is zero. Therefore,
\begin{equation}\label{eq:zero-orbit}
k \frac{|\sin(\psi)|}{(1-\sin(i)^2\sin(\psi)^2)^{\varepsilon '/3}} = (r/r_0 )^{-5/3} \left[ 1- \frac{V_0^2}{(r/r_0 )} \right].
\end{equation}
We also define the right-hand side of the above equation as $f(r/r_{0})$.
Since  radiation pressure vanishes at the equatorial plane, the above condition for the force-free line implies $V_0 = 1$. So, equation (\ref{eq:L-r0}) gives us
\begin{equation}
r_0 = \frac{L^2}{GM}.
\end{equation}

\begin{figure}
\includegraphics[scale=0.55]{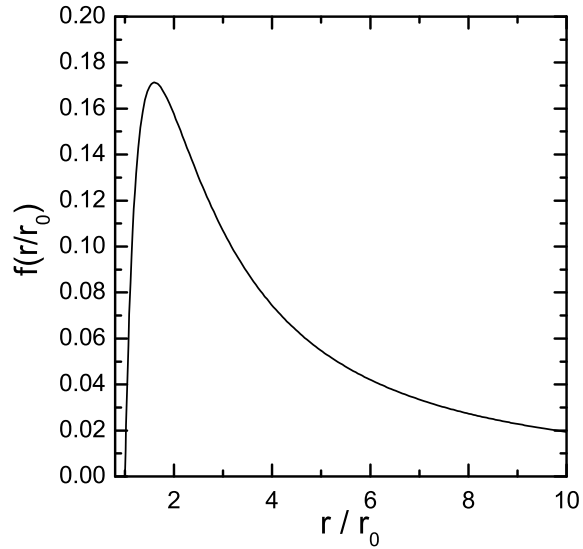}
\caption{Profile of $f(r/r_0 )$ (i.e., equation (\ref{eq:zero-orbit})) as a function of the normalized radial distance. Maximum value of this function is $0.17$.}
\label{fig:f2}
\end{figure}

The orbit is bound if the left-hand side of equation (\ref{eq:zero-orbit}) becomes less than the maximum value of the right-hand side of this equation (i.e., $f_{\rm max}$) for all values of $\psi$. According to Figure \ref{fig:f2} which shows profile of $f(r/r_0 )$, the maximum value of this function (i.e., the right-hand side of equation (\ref{eq:zero-orbit})) is $0.17$. Thus, as long as for all values of $\psi$ we have
\begin{equation}
k \frac{|\sin(\psi)|}{(1-\sin(i)^2\sin(\psi)^2)^{\varepsilon '/3}} < 0.17 ,
\end{equation}
then the orbit would be bound. Maximum value of the left hand side of the above inequality is $k/\cos(i)^{(2\varepsilon ' /3)}$. Thus, the requirement  of the existence of bound orbits becomes
\begin{equation}
k < 0.17 \cos(i)^{(2\varepsilon ' /3)} .
\end{equation}
Upon substituting from equation (\ref{eq:k}) into the above inequality, we obtain
\begin{equation}
N_{0} > N_{\rm cr}.
\end{equation}
where $N_{\rm cr}$ is the critical column density, i.e.
\begin{equation}
N_{\rm cr} = 17.6 \frac{ l}{\mu \sigma_{\rm T}} \sin(i)\cos(i)^{-2\varepsilon ' /3} [P_{0}(\varepsilon ', \alpha, \dot{m})]^{-2/3} .
\end{equation}

Thus, we obtained a  critical column density, above which bound orbits should be found. If we set $\varepsilon '=0$, this condition reduces to the critical density which has been found previously by \cite{krause11} (see their inequality (13)). But, here, our new condition depends on the properties of the gaseous component like the amount of the advected energy through the parameter $\varepsilon '$, the mass accretion rate $\dot{m}$ and the viscosity coefficient $\alpha$.  The above condition ensures the stability of cloud orbits given certain assumptions. But the contrary is not true as it has been mentioned by \cite{krause11}, i.e.  that
the cloud orbits must necessarily be unstable if the condition is not met. Here, we do not repeat the analysis in \cite{krause11}, but we note that orbits with unusually low column density also exist.

\begin{figure}
\includegraphics[scale=0.55]{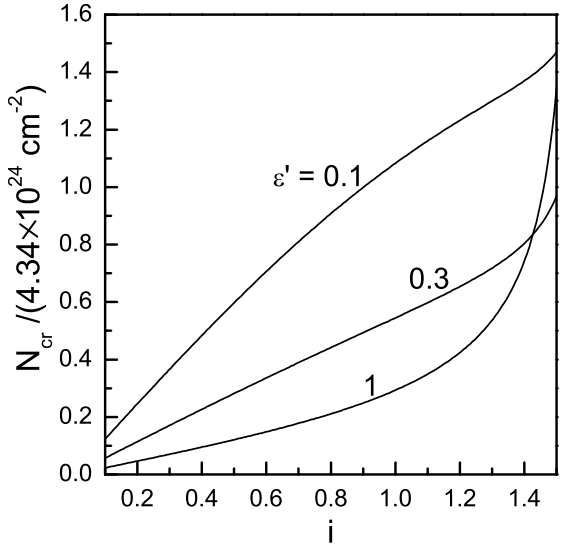}
\caption{Variation of the normalized critical column density, i.e. $N_{\rm cr}/(4.34\times 10^{24} {\rm cm}^{-2})$, as a function of the inclination angle $i$ for $\epsilon ' = 0.1$, $0.3$ and $1$. Beyond this critical column density bound orbits could be found.}
\label{fig:f3}
\end{figure}

Dependence of the critical density to the inclination angle $i$ is more complicated in comparison to the case considered by \cite{krause11}. Figure \ref{fig:f3} shows variations of the critical column density $N_{\rm cr}$ versus the inclination angle $i$ for different amounts of the advected energy. As the inclination angle increases, the critical column density increases irrespective of the value of $\epsilon'$. But this enhancement is more significant for the cases with a smaller value of the advected energy.  When the clouds are orbiting near to the equatorial plane, this critical column density has its minimum value.  Thus, population of clouds near to the equatorial region should be larger than other regions because a smaller value of the column density is needed to have long-lived stable clouds. On the other hand, it is unlikely to have clouds with a very large column density. Depending on the maximum value of the allowed column density, there would be a critical inclination angle $i_{\rm cr}$ beyond which clouds are {\it not} stable. Thus, there would be more clouds in the regions with the inclination angle less than $i_{\rm cr}$ and one may expect to observe more clouds near to the equatorial regions.

\begin{figure}
\includegraphics[scale=0.55]{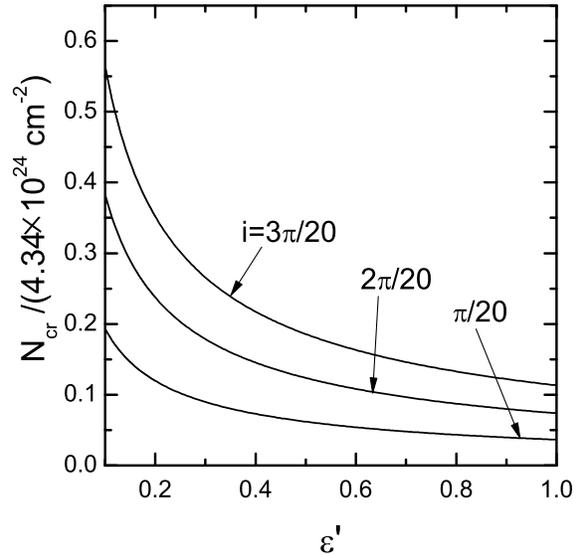}
\caption{Variation of the normalized critical column density, i.e. $N_{\rm cr}/(4.34\times 10^{24} {\rm cm}^{-2})$, as a function of $\epsilon'$ for the inclination angle $i=\pi/20$, $2\pi/20$ and $3\pi/20$.}
\label{fig:f4}
\end{figure}

Figure \ref{fig:f4} shows profile of $N_{\rm cr}$ as a function of $\epsilon'$ for different values of the inclination angle $i$. A larger inclination angle $i$ implies a larger critical column density. But as the intercloud medium becomes more advective, the critical column density reduces, though this reduction is not very significant for the clouds near to the equatorial region.

\begin{figure}
\includegraphics[scale=0.55]{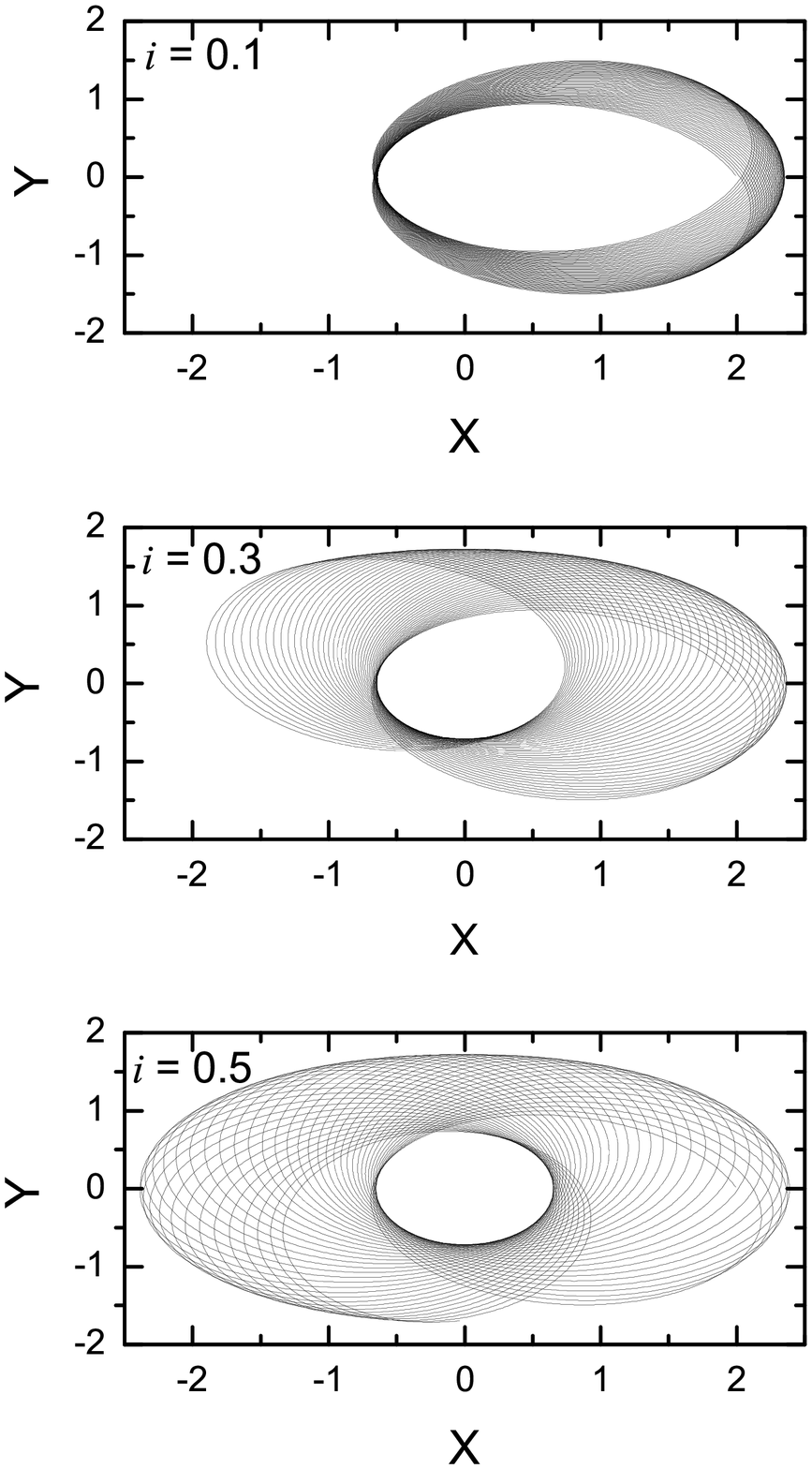}
\caption{Characteristic shapes of bound orbits in the orbital plane for $k_0 =0.1$, $\alpha = 0.1$, $\dot{m}=1.0$, $u(0)=0.5$ and $(du/d\psi)_{\psi =0}=0.2$ and different values of the inclination angle, i.e. $i=0.1$, $0.3$ and $0.5$.}
\label{fig:f5}
\end{figure}
\begin{figure}
\includegraphics[scale=0.55]{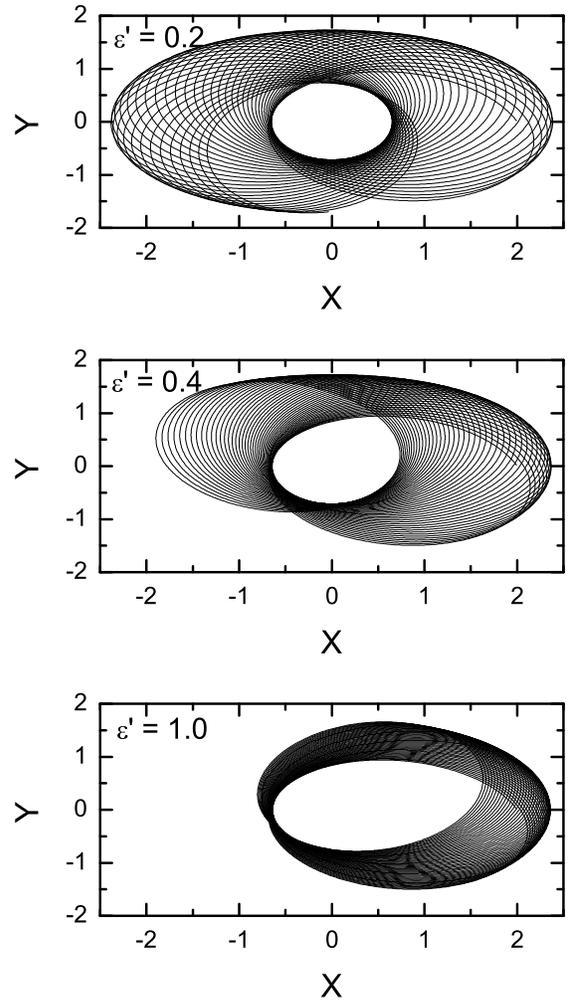}
\caption{The same as Figure \ref{fig:f5}, but for a fixed inclination angle $i=0.3$ and different values of the advected energy, i.e. $\epsilon '=0.2$, $0.4$ and $1.0$.}
\label{fig:f6}
\end{figure}

\subsection{SHAPE OF ORBITS}

To determine shape of the bound orbits, we plug into equation (\ref{eq:radial-r}) for $i$ and $k$, and numerically solve it subject to the initial conditions at $\psi =0$. But we note that parameter $k$ is obtained from the other input parameters according to equation (\ref{eq:k}). We explored bound orbits for various values of the initial values $u(\psi =0)$ and $(du/d\psi )_{\psi =0}$. But the effect of the input parameters $i$ and $k$ on the shape of orbits is qualitatively the same, irrespective of the initial values of $u$ and its derivative at $\psi =0$. Thus, we fixed the initial values as $u(0)=0.5$ and $(du/d\psi)_{\psi =0}=0.2$, but then the inclination angle $i$ and the parameter $\epsilon'$ are changed to study shape of the orbits. We also assume $k_0 =0.1$, $\alpha = 0.1$ and $\dot{m}=1.0$.
Figures \ref{fig:f5} and \ref{fig:f6} show shape of orbits for the above mentioned input parameters, but different values of the inclination angle $i$ and the advected energy $\epsilon'$.

Effect of the inclination angle $i$ is examined in Figure \ref{fig:f5}. The orbits look like   Rosetta orbits. All the three explored cases are for a fixed number of orbital rotation around the central mass. While orbits of the clouds with small inclination angles  are very close to each other, orbits of clouds with a larger inclination angle cover a larger volume of space. In other words, direction of cloud's motion changes much faster for the clouds with a larger inclination angle. A similar effect is seen in Figure \ref{fig:f6} where we keep the angle $i$  fixed,  but the advected energy is changed. As the gaseous component becomes less advective, then direction of the cloud's motion changes faster.

 In comparison to the analysis of \cite{krause11}, we only considered variations of intercloud's pressure with the polar angle based on a physical model. Despite of our little knowledge about the gaseous medium in which the clouds are moving, we prescribed it using ADAF model in the light of findings for the Galactic center and possibly other AGNs. In this regards, our analysis has more solid foundation comparing to the previous works where the intercloud medium is prescribed arbitrary. One of the key factors in constructing an ADAF model is the amount of the advected energy. We found that this parameter has a vital role in cloud's orbit analysis. Moreover, ADAFs are geometrically thick and each physical variables has a polar angle dependence. Exactly because of such a dependence, our condition for the existence of bound orbits differs from a case without considering polar angle's dependence which has been explored in \cite{krause11}.

\section{Conclusions}
We studied orbital motion of BLR clouds considering forces due to the gravity and a non-isotropic radiation source. Previous related studies are generalized by prescribing pressure profile of the intercloud medium as a function of the radial distance and the latitudinal angle.  This prescription is based on an analytical solution for ADAFs which has been reported just recently by Shadmehri (2014). Our stability analysis for our spherical, pressure-confined clouds implies that the spatial distribution of the clouds tends to populate the equatorial region more than other regions, and this finding is independent of the true nature of the intercloud medium provided the pressure profile of the ambient medium increases from the pole to the equatorial regions and the clouds are pressure-confined. In other words, a disc-like configuration is more plausible for the distribution of the BLR clouds.  We emphasize that previous studies  \citep[e.g.,][]{krause11} reached to the  conclusion that BLRs should be more disc-like from the stability point of view and this argument is significantly corroborated by the present analysis.

\section*{Acknowledgments}

I am grateful to the referee,    for helpful comments and suggestions that improved the paper.

\bibliographystyle{mn2e}
\bibliography{reference}

\begin{thebibliography}{}

\bibitem[\protect\citeauthoryear{{Burkert}, {Schartmann}, {Alig}, {Gillessen},
  {Genzel}, {Fritz} \& {Eisenhauer}}{{Burkert} et~al.}{2012}]{burkert2012}
{Burkert} A.,  {Schartmann} M.,  {Alig} C.,  {Gillessen} S.,  {Genzel} R.,
  {Fritz} T.~K.,    {Eisenhauer} F.,  2012, ApJ, 750, 58

\bibitem[\protect\citeauthoryear{{Gillessen}, {Genzel}, {Fritz}, {Quataert},
  {Alig}, {Burkert}, {Cuadra}, {Eisenhauer}, {Pfuhl}, {Dodds-Eden}, {Gammie} \&
  {Ott}}{{Gillessen} et~al.}{2012}]{Gill}
{Gillessen} S.,  {Genzel} R.,  {Fritz} T.~K.,  {Quataert} E.,  {Alig} C.,
  {Burkert} A.,  {Cuadra} J.,  {Eisenhauer} F.,  {Pfuhl} O.,  {Dodds-Eden} K.,
  {Gammie} C.~F.,    {Ott} T.,  2012, Nature, 481, 51

\bibitem[\protect\citeauthoryear{{Khajenabi}, {Rahmani} \&
  {Abbassi}}{{Khajenabi} et~al.}{2014}]{khajenabi2014}
{Khajenabi} F.,  {Rahmani} M.,    {Abbassi} S.,  2014, MNRAS, 439, 2468

\bibitem[\protect\citeauthoryear{{Krause}, {Burkert} \& {Schartmann}}{{Krause}
  et~al.}{2011}]{krause11}
{Krause} M.,  {Burkert} A.,    {Schartmann} M.,  2011, MNRAS, 411, 550

\bibitem[\protect\citeauthoryear{{Krause}, {Schartmann} \& {Burkert}}{{Krause}
  et~al.}{2012}]{krause12}
{Krause} M.,  {Schartmann} M.,    {Burkert} A.,  2012, MNRAS, 425, 3172

\bibitem[\protect\citeauthoryear{{Krolik} \& {Begelman}}{{Krolik} \&
  {Begelman}}{1988}]{krolik88}
{Krolik} J.~H.,  {Begelman} M.~C.,  1988, ApJ, 329, 702

\bibitem[\protect\citeauthoryear{{Liu} \& {Zhang}}{{Liu} \&
  {Zhang}}{2011}]{Liu2012}
{Liu} Y.,  {Zhang} S.~N.,  2011, ApJL, 728, L44

\bibitem[\protect\citeauthoryear{{Maiolino}, {Risaliti}, {Salvati}, {Pietrini},
  {Torricelli-Ciamponi}, {Elvis}, {Fabbiano}, {Braito} \& {Reeves}}{{Maiolino}
  et~al.}{2010}]{maiolino2010}
{Maiolino} R.,  {Risaliti} G.,  {Salvati} M.,  {Pietrini} P.,
  {Torricelli-Ciamponi} G.,  {Elvis} M.,  {Fabbiano} G.,  {Braito} V.,
  {Reeves} J.,  2010, A\& A, 517, A47

\bibitem[\protect\citeauthoryear{{Marconi}, {Axon}, {Maiolino}, {Nagao},
  {Pastorini}, {Pietrini}, {Robinson} \& {Torricelli}}{{Marconi}
  et~al.}{2008}]{Marco}
{Marconi} A.,  {Axon} D.~J.,  {Maiolino} R.,  {Nagao} T.,  {Pastorini} G.,
  {Pietrini} P.,  {Robinson} A.,    {Torricelli} G.,  2008, ApJ, 678, 693

\bibitem[\protect\citeauthoryear{{Mioc} \& {Radu}}{{Mioc} \&
  {Radu}}{1992}]{Mio}
{Mioc} V.,  {Radu} E.,  1992, Astronomische Nachrichten, 313, 353

\bibitem[\protect\citeauthoryear{{Namekata}, {Umemura} \&
  {Hasegawa}}{{Namekata} et~al.}{2014}]{nameka}
{Namekata} D.,  {Umemura} M.,    {Hasegawa} K.,  2014, MNRAS, 443, 2018

\bibitem[\protect\citeauthoryear{{Narayan} \& {Yi}}{{Narayan} \&
  {Yi}}{1994}]{Narayan94}
{Narayan} R.,  {Yi} I.,  1994, ApJL, 428, L13

\bibitem[\protect\citeauthoryear{{Narayan} \& {Yi}}{{Narayan} \&
  {Yi}}{1995}]{Narayan95}
{Narayan} R.,  {Yi} I.,  1995, ApJ, 444, 231

\bibitem[\protect\citeauthoryear{{Nenkova}, {Ivezi{\'c}} \&
  {Elitzur}}{{Nenkova} et~al.}{2002}]{nenkova}
{Nenkova} M.,  {Ivezi{\'c}} {\v Z}.,    {Elitzur} M.,  2002, ApJL, 570, L9

\bibitem[\protect\citeauthoryear{{Netzer}}{{Netzer}}{2013}]{netzerbook}
{Netzer} H.,  2013, {The Physics and Evolution of Active Galactic Nuclei,
  Cambridge, UK: Cambridge University Press, 2013}

\bibitem[\protect\citeauthoryear{{Netzer} \& {Marziani}}{{Netzer} \&
  {Marziani}}{2010}]{netzer10}
{Netzer} H.,  {Marziani} P.,  2010, ApJ, 724, 318

\bibitem[\protect\citeauthoryear{{Plewa}, {Schartmann} \& {Burkert}}{{Plewa}
  et~al.}{2013}]{plewa}
{Plewa} P.~M.,  {Schartmann} M.,    {Burkert} A.,  2013, MNRAS, 431, L127

\bibitem[\protect\citeauthoryear{{Proga}, {Jiang}, {Davis}, {Stone} \&
  {Smith}}{{Proga} et~al.}{2014}]{proga}
{Proga} D.,  {Jiang} Y.-F.,  {Davis} S.~W.,  {Stone} J.~M.,    {Smith} D.,
  2014, ApJ, 780, 51

\bibitem[\protect\citeauthoryear{{Rees}}{{Rees}}{1987}]{Ress}
{Rees} M.~J.,  1987, MNRAS, 228, 47P

\bibitem[\protect\citeauthoryear{{Risaliti}, {Nardini}, {Salvati}, {Elvis},
  {Fabbiano}, {Maiolino}, {Pietrini} \& {Torricelli-Ciamponi}}{{Risaliti}
  et~al.}{2011}]{risaliti2011}
{Risaliti} G.,  {Nardini} E.,  {Salvati} M.,  {Elvis} M.,  {Fabbiano} G.,
  {Maiolino} R.,  {Pietrini} P.,    {Torricelli-Ciamponi} G.,  2011, MNRAS,
  410, 1027

\bibitem[\protect\citeauthoryear{{Saslaw}}{{Saslaw}}{1978}]{saslaw}
{Saslaw} W.~C.,  1978, ApJ, 226, 240

\bibitem[\protect\citeauthoryear{{Schartmann}, {Burkert}, {Alig}, {Gillessen},
  {Genzel}, {Eisenhauer} \& {Fritz}}{{Schartmann} et~al.}{2012}]{schart}
{Schartmann} M.,  {Burkert} A.,  {Alig} C.,  {Gillessen} S.,  {Genzel} R.,
  {Eisenhauer} F.,    {Fritz} T.~K.,  2012, ApJ, 755, 155

\bibitem[\protect\citeauthoryear{{Schartmann}, {Meisenheimer}, {Camenzind},
  {Wolf}, {Tristram} \& {Henning}}{{Schartmann} et~al.}{2008}]{schartmann2008}
{Schartmann} M.,  {Meisenheimer} K.,  {Camenzind} M.,  {Wolf} S.,  {Tristram}
  K.~R.~W.,    {Henning} T.,  2008, A\& A, 482, 67

\bibitem[\protect\citeauthoryear{{Shadmehri}}{{Shadmehri}}{2014}]{shadmehri14}
{Shadmehri} M.,  2014, MNRAS, 442, 3528

\bibitem[\protect\citeauthoryear{{Shakura} \& {Sunyaev}}{{Shakura} \&
  {Sunyaev}}{1973}]{shakura}
{Shakura} N.~I.,  {Sunyaev} R.~A.,  1973, A\& A, 24, 337

\bibitem[\protect\citeauthoryear{{Torricelli-Ciamponi}, {Pietrini}, {Risaliti}
  \& {Salvati}}{{Torricelli-Ciamponi} et~al.}{2014}]{torri}
{Torricelli-Ciamponi} G.,  {Pietrini} P.,  {Risaliti} G.,    {Salvati} M.,
  2014, MNRAS, 442, 2116

\bibitem[\protect\citeauthoryear{{Wang}, {Cheng} \& {Li}}{{Wang}
  et~al.}{2012}]{wang2012}
{Wang} J.-M.,  {Cheng} C.,    {Li} Y.-R.,  2012, ApJ, 748, 147

\end{thebibliography}

\end{document}